\documentstyle[aps,prl,twocolumn,epsf]{revtex}

\begin{document}
\draft
\title{Impurity-Induced Antiferromagnetic Ordering in the Spin Gap System TlCuCl$_3$}

\author{Akira Oosawa, Toshio Ono and Hidekazu Tanaka}

\address{
Department of Physics, Tokyo Institute of Technology,
Oh-okayama, Meguro-ku, Tokyo 152-8551, Japan
}

\date{\today}

\maketitle

\begin{abstract}
The magnetization measurements have been performed on the doped spin gap system TlCu$_{1-x}$Mg$_x$Cl$_3$ with $x\leq 0.025$. The parent compound TlCuCl$_3$ is a three-dimensional coupled spin dimer system with the excitation gap ${\Delta}/k_{\rm B}=7.7$ K. The impurity-induced antiferromagnetic ordering was clearly observed. The easy axis lies in the $(0,1,0)$ plane. It was found that the transition temperature increases with increasing Mg$^{2+}$ concentration $x$, while the spin-flop transition field is almost independent of $x$. The magnetization curve suggests that the impurity-induced antiferromagnetic ordering coexists with the spin gap for $x \leq 0.017$. 
\end{abstract}


The singlet ground state with an excitation gap (spin gap) has been found in many quantum spin systems such as the Haldane chain, spin-Peierls system, spin ladder, exchange alternating chain, and coupled spin dimer system. These spin gap systems do not undergo magnetic ordering down to zero temperature. In the last decade, however, it has been found that very small amounts of nonmagnetic site impurities can give rise to antiferromagnetic ordering in these spin gap systems \cite{Hase,Oseroff,Masuda,Martin,Azuma,Uchiyama}. When nonmagnetic ions are substituted for magnetic ions, the singlet ground state is disturbed so that staggered moments are induced around the impurities. If the induced moments interact through effective exchange interactions, which are mediated by intermediate singlet spins, three-dimensional (3D) long-range order can arise. Such impurity-induced magnetic ordering is a new type of phase transition caused by the quantum effect.

Impurity-induced magnetic ordering of spin gap systems was first observed in Zn$^{2+}$-doped CuGeO$_3$ \cite{Hase}, which is a well-known inorganic spin-Peierls material \cite{Hase2}, and was precisely investigated in Mg$^{2+}$-doped CuGeO$_3$ \cite{Masuda}. Magnetic ordering due to pinning of lattice dimerization by doped nonmagnetic impurities was also observed in CuGe$_{1-x}$Si$_x$O$_3$ \cite{Regnault}. In Cu$_{1-x}$Zn$_x$GeO$_3$ and CuGe$_{1-x}$Si$_x$O$_3$ with low $x$, the coexistence of lattice dimerization and antiferromagnetic ordering has been observed in neutron scattering experiments \cite{Martin,Regnault}. Impurity-induced antiferromagnetic ordering was also observed in Zn$^{2+}$-doped antiferromagnetic two-leg spin ladder SrCu$_2$O$_3$ \cite{Azuma} and the Mg$^{2+}$-doped Haldane system PbNi$_2$V$_2$O$_8$ \cite{Uchiyama}. 
The site-impurity effect on spin gap systems has been argued theoretically by many authors \cite{Fukuyama,Miyazaki,Yasuda}, and impurity-induced antiferromagnetic ordering has been demonstrated. 

As mentioned above, experimental studies of impurity-induced antiferromagnetic ordering have been performed only on quasi-1D systems to date. Therefore, for comprehensive understanding of impurity-induced antiferromagnetic ordering, it may be necessary to study the impurity effect on 3D spin gap systems. With this motivation, we studied the magnetic properties in the Mg$^{2+}$-doped spin gap system TlCuCl$_3$ by means of magnetization measurements. 

TlCuCl$_3$ has the monoclinic structure (space group $P2_1/c$) \cite{Takatsu}. The crystal structure is composed of planar dimers of Cu$_2$Cl$_6$. The dimers are stacked on top of one another and form infinite double chains parallel to the crystallographic $a$-axis. These double chains are located at the corners and center of the unit cell in the $b-c$ plane, and are separated by Tl$^+$ ions. The magnetic ground state of TlCuCl$_3$ is the spin singlet with excitation gap $\Delta=7.7$ K \cite{Takatsu,Shiramura,Oosawamag}. From the analyses of the dispersion relation obtained by neutron inelastic scattering, it was found that the origin of the spin gap of TlCuCl$_3$ is the strong antiferromagnetic interaction $J=5.68$ meV in the chemical dimer Cu$_2$Cl$_6$, and that the neighboring dimers are coupled by the strong interdimer interactions along the double chain and in the $(1, 0, -2)$ plane \cite{Cavadini,Oosawaneu}. Consequently, TlCuCl$_3$ was characterized as a strongly coupled 3D spin dimer system. 

Before preparing the doped TlCu$_{1-x}$Mg$_x$Cl$_3$ system, we prepared single crystals of TlCuCl$_3$. The preparation of a single crystal of TlCuCl$_3$ is described in reference \cite{Oosawamag}. Mixing TlCuCl$_3$, TlCl and MgCl$_2$ in a ratio of $1-x : x : x$, we prepared TlCu$_{1-x}$Mg$_x$Cl$_3$ by the vertical Bridgman method. We obtained single crystals of 1$\sim$3 cm$^3$ with $x=0.008,\ 0.014,\ 0.017,\ 0.020,\ 0.022$ and 0.025. The magnesium concentration $x$ was analyzed by emission spectrochemical analysis after the measurements. Samples used for magnetic measurements were cut to $50 \sim 150$ mg. The magnetizations were measured down to 1.8 K in magnetic fields up to 7 T using a SQUID magnetometer (Quantum Design MPMS XL). 
 
\begin{figure}[ht]
\epsfxsize=70mm
\centerline{\epsfbox{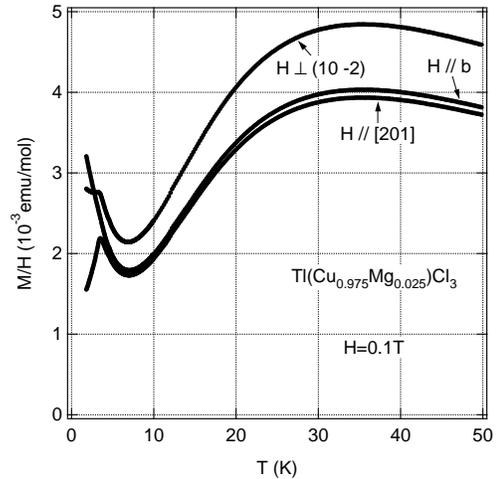}}
\caption{Temperature dependence of the magnetic susceptibilities $\chi=M/H$ in TlCu$_{0.975}$Mg$_{0.025}$Cl$_3$ measured at $H=0.1$ T for $H\parallel b$, $H\perp (1,0,{\bar 2})$ and $H\parallel [2,0,1]$. }
\label{fig1}
\end{figure}

First, we measured the temperature dependence of the magnetic susceptibilities $\chi=M/H$ in TlCu$_{0.975}$Mg$_{0.025}$Cl$_3$ under the magnetic field of $H=0.1$ T perpendicular to the cleavage planes  $(0,1,0)$ and $(1,0,{\bar 2})$, and along the $[2,0,1]$ direction which is parallel to both $(0,1,0)$ and $(1,0,{\bar 2})$ planes, respectively. These three field directions are orthogonal to one another. The results are shown in Fig. \ref{fig1}. With decreasing temperature, the magnetic susceptibilities display broad maxima at $T\sim 36$ K and decrease, as observed in pure TlCuCl$_3$ \cite{Takatsu,Oosawamag}. However, the magnetic susceptibilities increase again below 7 K, and then exhibit sharp bend anomalies at $T_{\rm N}=3.5$ K, which is indicative of magnetic ordering, for $H \perp (1,0,{\bar 2})$ and $H \parallel [2,0,1]$, whereas the magnetic susceptibility for $H \parallel b$ increases monotonically. Below $T_{\rm N}$, the magnetic susceptibility for $H \perp (1,0,{\bar 2})$ is almost independent of temperature, while that for $H \parallel [2,0,1]$ decreases. These results indicate that the antiferromagnetic ordering induced by Mg$^{2+}$ doping occurs at $T_{\rm N}=3.5$ K, and that the easy axis is in the $(0,1,0)$ plane and is close to the $[2,0,1]$ direction. 

In order to determine the easy axis, we measured the magnetic susceptibility in TlCu$_{0.980}$Mg$_{0.020}$Cl$_3$, varying the field direction in the $(0,1,0)$ plane at $T=1.8$ K. The susceptibility minimum was observed, when the magnetic field was at an angle of 38$^{\circ}$ with the $a$-axis. This implies that the easy axis is canted by 38$^{\circ}$ from the $a$-axis. Since the angle between the $[2,0,1]$ direction and the $a$-axis is 51$^{\circ}$, the angle between the $[2,0,1]$ direction and the easy axis is 13$^{\circ}$. 

\begin{figure}[ht]
\epsfxsize=70mm
\centerline{\epsfbox{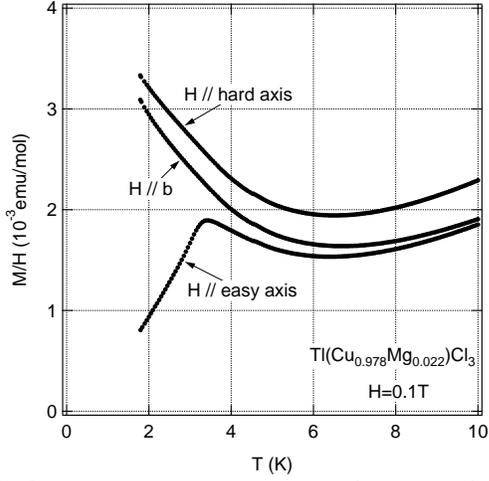}}
\caption{Low-temperature magnetic susceptibilities in TlCu$_{0.978}$Mg$_{0.022}$Cl$_3$ measured at $H=0.1$ T for $H\parallel$ easy axis, $H\parallel$ hard axis in the $(0,1,0)$ plane and $H\parallel b$.}
\label{fig2}
\end{figure}

Figure \ref{fig2} shows the low-temperature magnetic susceptibilities in TlCu$_{0.978}$Mg$_{0.022}$Cl$_3$ measured at $H=0.1$ T. The magnetic field was applied along the $b$-axis, the easy axis and the hard axis in the $(0,1,0)$ plane, which are orthogonal to one another. With decreasing temperature, the magnetic susceptibility of the easy axis decreases toward zero below $T_{\rm N}=3.45$ K, while the others increase monotonically. This behavior is indicative of an antiferromagnet with collinear ordering.

\begin{figure}[ht]
\epsfxsize=70mm
\centerline{\epsfbox{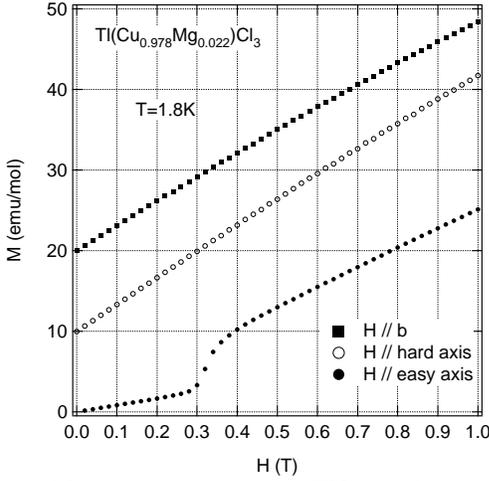}}
\caption{Magnetization curves in TlCu$_{0.978}$Mg$_{0.022}$Cl$_3$ measured at $T=1.8$ K for $H\parallel$ easy axis, $H\parallel$ hard axis in the $(0,1,0)$ plane and $H\parallel b$. The values of magnetization are shifted upward by 10 emu/mol.}
\label{fig3}
\end{figure}

Figure \ref{fig3} shows the magnetization curves of TlCu$_{0.978}$Mg$_{0.022}$Cl$_3$ measured at $T=1.8$ K for $H \parallel$ easy axis, hard axis and $b$-axis. For $H \parallel$ easy axis, the spin-flop transition is clearly observed at $H_{\rm sp}=0.32$ T, while no transition is observed for the other field directions. From these results, it is evident that collinear antiferromagnetic ordering along the easy axis arises below $T_{\rm N}$. The sharp cusplike maximum of the susceptibility at $T_{\rm N}$ and the sharp spin-flop transition indicate the good homogeneity of the present sample at the macroscopic scale.

\begin{figure}[ht]
\epsfxsize=70mm
\centerline{\epsfbox{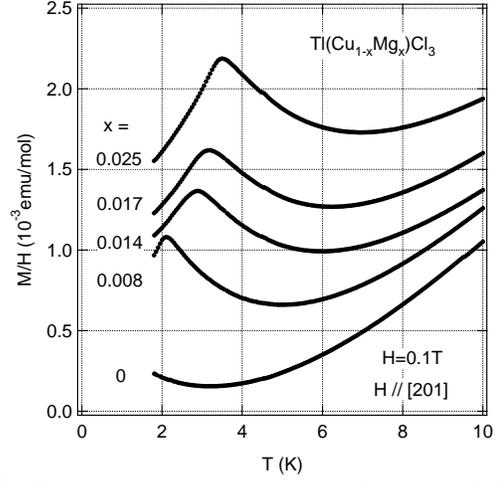}}
\caption{Low-temperature magnetic susceptibilities in TlCu$_{1-x}$Mg$_{x}$Cl$_3$ measured at $H=0.1$ T for $H \parallel [2,0,1]$ for various $x$.}
\label{fig4}
\end{figure}

\begin{figure}[ht]
\epsfxsize=70mm
\centerline{\epsfbox{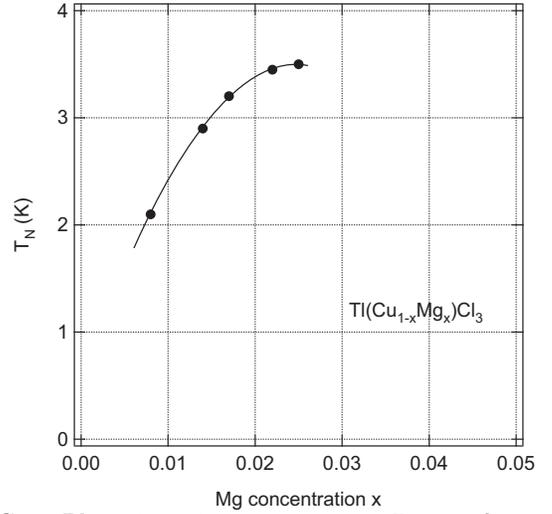}}
\caption{Phase transition temperature $T_{\rm N}$ as a function of $x$ in TlCu$_{1-x}$Mg$_{x}$Cl$_3$. The solid line is a guide for the eyes.}
\label{fig5}
\end{figure}

Figure \ref{fig4} shows the temperature dependence of the magnetic susceptibilities in TlCu$_{1-x}$Mg$_{x}$Cl$_3$ with various $x$ for $H \parallel [2,0,1]$. The magnetic field of 0.1 T was applied in the measurements. For $x=0$, no anomaly indicative of the magnetic ordering can be seen down to 1.8 K, although a slight increase of the magnetic susceptibility was observed due to a small amount of impurities or lattice defects. On the other hand, the cusplike anomalies indicative of antiferromagnetic ordering were clearly observed for $x \neq 0$. With increasing $x$, the transition temperature $T_{\rm N}$ and the magnitude of the magnetic susceptibility increase. The transition temperatures for $x\leq 0.025$ were plotted in Fig. \ref{fig5} as a function of $x$. The transition temperature $T_{\rm N}$ shows a tendency to saturate at $T_{\rm N}\approx 3.5$ K. Although the behavior of the phase transition for $x > 0.025$ is of great interest, the measurements were not performed, because it is difficult to prepare good single crystals of TlCu$_{1-x}$Mg$_{x}$Cl$_3$ with $x > 0.025$ at present. 
 
\begin{figure}[ht]
\epsfxsize=70mm
\centerline{\epsfbox{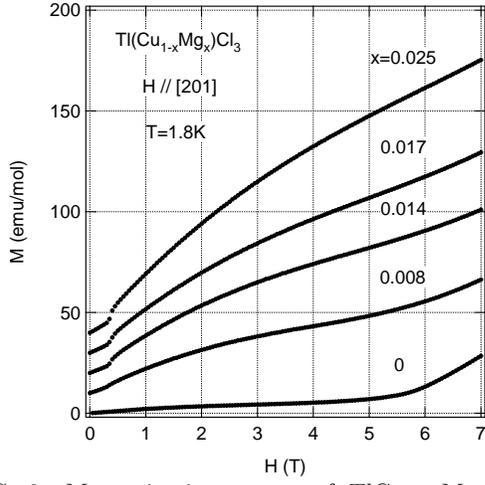}}
\caption{Magnetization curves of TlCu$_{1-x}$Mg$_{x}$Cl$_3$ measured at $T=1.8$ K for $H \parallel [2,0,1]$ for various $x$. The values of magnetization are shifted upward by 10 emu/mol with increasing $x$.}
\label{fig6}
\end{figure}

It is considered that the direction of the easy axis does not strongly depend on $x$, and is close to the $[2,0,1]$ direction, because the susceptibilities for various $x$ display a similar cusplike anomaly for $H\parallel [2,0,1]$, as shown in Fig. \ref{fig4}. Thus, it is expected that the spin-flop transition is observed for $H\parallel [2,0,1]$, and the transition field is close to that for $H \parallel$ easy axis. Figure \ref{fig6} shows the magnetization curves of TlCu$_{1-x}$Mg$_{x}$Cl$_3$ with various $x$ measured at $T=1.8$ K for $H \parallel [2,0,1]$. For $x=0$, the magnetization is almost zero up to the gap field $H_{\rm g}=\Delta/g\mu_{\rm B} \approx 6$ T, and then it increases rapidly. For $x \neq 0$, the magnetization increases with a finite slope, and exhibits a jump indicative of the spin-flop transition at $H \approx 0.35$ T with increasing magnetic field. It is noted that spin-flop field $H_{\rm sp}$ is almost independent of $x$, although the amount of the magnetization jump increases with increasing $x$.

A theoretical description of the spin-flop transition in the impurity-induced antiferromagnetic phase has not been established to date. In the conventional antiferromagnet, spin-flop field $H_{\rm sp}$ is proportional to the square root of the product of anisotropy energy and the exchange interactions and the magnitude of the spin moment. The small value of $\approx 0.35$ T of spin-flop field $H_{\rm sp}$ may be indicative of the low induced spin moments around the impurities or the low anisotropy energy. At present, however, we have no explanation for why the spin-flop field is independent of $x$.

For $x \leq 0.017$, the magnetization tends to increase at $H \sim 6$ T, which is almost the same as the $\approx 6$ T of gap field $H_{\rm g}$ of pure TlCuCl$_3$, as shown in Fig. \ref{fig6}. This suggests that the excitation gap remains for $x \leq 0.017$, {\it i.e.}, antiferromagnetic ordering coexists with the excitation gap. If the gap remains, field-induced magnetic ordering may occur, as observed in pure TlCuCl$_3$ \cite{Oosawamag,Oosawaheat,Tanakaela}. In TlCuCl$_3$, the magnetization has the cusplike minimum at the transition temperature \cite{Oosawamag,Oosawaheat}. This behavior can be described in terms of the Bose-Einstein condensation of the excited triplets (magnons) \cite{Nikuni}. In the ordered phase, transverse staggered magnetic ordering with respect to the applied magnetic field occurs \cite{Nikuni,Tachiki1,Tachiki2}. This transverse staggered magnetic ordering, in which two spins on the same dimer are antiparallel, was confirmed by neutron elastic scattering analysis in high magnetic fields for $H\parallel b$ \cite{Tanakaela}. For this field direction, it was observed that spins lie in the (0,1,0) plane at an angle of 39$^{\circ}$ to the $a$-axis. This spin direction is almost the same as that of the easy axis in the impurity-induced antiferromagnetic phase for TlCu$_{0.978}$Mg$_{0.022}$Cl$_3$. Thus, the relation between the impurity-induced antiferromagnetic phase and the field-induced ordered phase is of great interest. 

In order to investigate whether or not field-induced magnetic ordering occurs in Mg$^{2+}$-doped TlCuCl$_3$, we examined the temperature dependence of the magnetization in TlCu$_{0.992}$Mg$_{0.008}$Cl$_3$ at various magnetic fields up to 7 T for $H \parallel [2,0,1]$. However, no definite phase transition was detected due to the large Curie-Weiss term, although a fairly sharp inflection point indicative of the phase transition was observed for $H>5$ T. Thus, other experiments such as specific heat and neutron scattering measurement are required in order to determine whether field-induced magnetic ordering occurs also in Mg$^{2+}$-doped TlCuCl$_3$. 

In conclusion, we carried out magnetization measurements in TlCu$_{1-x}$Mg$_x$Cl$_3$. Impurity-induced antiferromagnetic ordering was clearly observed for $0.008 \le x \le 0.025$. It was found that the easy axis of the magnetic moments lies in the (0,1,0) plane at an angle of $38^{\circ}$ to the $a$-axis. For $x \le 0.025$, the transition temperature increases with increasing $x$, while the spin-flop field is almost independent of $x$. The magnetization curve suggests that the energy gap remains even in the ordered phase for $x \leq 0.017$. 

We thank T. Masuda for useful comments on impurity-induced antiferromagnetic ordering. This work was supported by the Toray Science Foundation and a Grant-in-Aid for Scientific Research on Priority Areas (B) from the Ministry of Education, Culture, Sports, Science and Technology of Japan. A. O. was supported by Research Fellowships of the Japan Society for the Promotion of Science for Young Scientists.

\end{document}